\documentclass[final,12pt,3p]{elsarticle}

\usepackage{amsmath,amssymb}
\usepackage{caption} 
\usepackage{makecell}
\usepackage[table]{xcolor}
\definecolor{Gray}{gray}{0.9}
\usepackage{multirow}
\usepackage{subcaption}
\makeatletter
\newcommand*{\rom}[1]{\expandafter\@slowromancap\romannumeral #1@}
\makeatother
\usepackage{amsthm}

\usepackage{hyperref}

\journal{International Journal of Electrical Power and Energy Systems}

\begin{document}

\begin{frontmatter}

\title{An Integrated Algorithm for Evaluating Plug-in Electric Vehicle's Impact on the State of Power Grid Assets}

\author[label1]{Daijiafan Mao\corref{cor1}}
\address[label1]{Department of Electrical and Computer Engineering, The Ohio State University, Columbus, OH 43210, USA}

\cortext[cor1]{Corresponding author at: 205 Dreese Labs, 2015 Neil Ave., Columbus, OH 43210, USA}

\ead{mao.156@osu.edu}

\author[label1]{Ziran Gao}
\ead{gao.1233@osu.edu}

\author[label1]{Jiankang Wang}
\ead{wang.6536@osu.edu}

\begin{abstract}
Plug-in Electric Vehicles (PEV) exert an increasingly disruptive influence on power delivery systems with penetration surge in the past decade. Therefore, accurately assessing their impact plays a crucial role in managing grid assets and maintaining power grids' reliability. However, PEV loads are stochastic and \emph{impulsive}, which means they are of high power density and vary in a fast and discrete manner. These load characteristics make conventional assessment methods unsuitable. This paper proposes an algorithm, which captures the inter-temporal response of grid assets and allows fast assessment through an integrated interface. To realize these advantageous features, we establish analytical models for two generic classes of grid assets (continuous and discrete operating assets) and recast their cost functions in the statistical settings of PEV charging. Distinct from simulation-based methods, the proposed method is analytical, and thus greatly reduce the computation resources and data required for accurate assessment. The effectiveness of the proposed algorithm has been demonstrated on a set of power distribution networks in Columbus metropolitan area, in comparison with the conventional assessment methods.
\end{abstract}


\end{frontmatter}




\section{Introduction}
\label{sec1}
The current electric power system has been increasingly penetrated with Plug-in Electric Vehicles (PEV). According to the International Energy Agency (IEA), over 750 thousand fleets of new PEVs were registered in 2016 alone, and the worldwide PEV penetration target is 30\% of total market share by 2030 \cite{IEAEVOutlook2017}. The power required to charge PEVs is provided at the distribution and potentially sub-transmission level (below 69 kV) of the grid \cite{meyer2018integrating}. PEV loads consume much higher power during charging. As Table~\ref{t1} shows, at DC Level 2, it is possible to charge a 25 kWh battery pack in 10 minutes, which far exceeds the peak power demand for an average household in the U.S. Moreover, the power electronics-interfaced (PE-interfaced) configuration of PEV charger can ramp to full charging level almost instantaneously. For example, it only takes 7 seconds for a 2016 Ford Focus Electric to reach its full charging power after connecting to the grid.
\begin{table}[h!]
\begin{center}
\caption {PEV Charging Ratings and Configurations \cite{IEAEVOutlook2017,yilmaz2013reviewbattery,shareef2016review,sae2011sae}} \label{t1} 
    \begin{tabular}{ | c | c | c | c | c | c |}
    \hline
    Charging Level & \thead{Input Voltage \\ and Connection} & \thead{Maximum Power \\ (kW)} & Charging Time & Typical Use \\ \hline
    AC Level 1 & \thead{120V \\ 1-phase} & 2 & 10$\sim$13h & \multirow{3}{*}{\thead{Private/Public \\ Residential/Commercial}} \\ \cline{1-4}
    AC Level 2 & \thead{240V \\ 1-phase/3-phase} & 20 & 1$\sim$4h  &\\ \cline{1-4}
    AC Level 3 & \thead{240V \\ 3-phase} & 43.5 & $\sim$1h & \\ \hline
    DC Level 1 & \thead{200$\sim$450V \\ 3-phase} & 36 & 0.5$\sim$1.44h & \multirow{3}{*}{\thead{Public \\ Commercial}} \\ \cline{1-4}
    DC Level 2 & \thead{200$\sim$450V \\ 3-phase} & 96 & 0.2$\sim$0.58h &  \\ \cline{1-4}
    DC Level 3 & \thead{200$\sim$600V \\ 3-phase} & 200 & $\sim$10min & \\ \hline
    \multicolumn{5}{c}{Note: AC Level 3 and DC Level 3 are not yet finalized }\\
    \end{tabular}
    \vspace{-20pt}
\end{center}
\end{table}

Distinct from conventional loads, PEV loads are stochastic and \emph{impulsive}, which means they are of high power density and vary in a fast and discrete manner. Prior works have shown that these load characteristics will result in negative impacts on the power grid, including disruptively varying voltage profiles along the feeder and overloading of service transformers \cite{shareef2016review,mu2014spatial,sun2016optimal,mao2017evaluating,gong2012study,yilmaz2013reviewimpact}. This will consequently affect the operating state of grid asset and induce asset depreciation over the long term. With increasing PEV penetration and improving fast/ultra-fast charging technologies, it is critical for electric utilities to accurately quantify the impact of PEV loads on grid assets and plan for equipment replacement and infrastructure expansion accordingly, in order to ensure service reliability.

On assessing grid assets' response under high penetration of PEVs, existing studies fall into two categories: static analysis and Time-Series (TS) analysis. Most of the static analysis results in the consideration of maximum PEV loads induced by coincidental charging. For example, \cite{fernandez2011assessment} shows that the energy losses can increase up to 40\% in off-peak hours and the investment cost can increase up to 15\% of total distribution network costs for a scenario of 60\% PEV penetration level. In \cite{shafiee2013investigating}, the case study shows that both peak-to-average ratio (PAR) and loss increment are the big concern to the widespread use of PEVs due to the coincidence of daily peak load and charging activities. The shortfall of this approach is that only the worst cases are considered, and thus tend to overestimate the PEV's impact. Improving on this approach, other work, such as \cite{leou2014stochastic,qian2011modeling}, considers the probabilistic distribution of PEV loads connected in the system. In \cite{leou2014stochastic}, Roulette wheel selection concept is used to take various uncertainties into account, thus quantifies the congestion and security risk impact of PEV in the form of probabilistic distribution functions. While these assessments allow more accurate input of PEV charging, an inherent deficiency of the static analysis is embedded from the assumption of fixed grid configurations. Therefore, they cannot capture the inter-temporal response of grid assets. These deficiencies can be alleviated in TS analysis. 

TS analysis feeds load profiles in time series to power flow analysis and observes power grid's response. A few studies adopt TS analysis in PEV's impact evaluation, under deterministic or stochastic settings. Ref. \cite{qian2011modeling}  simulated four PEV charging scenarios, considering stochastic nature in charging start time, and thus concludes that a 20\% level of PEV penetration would lead to a 35.8\% increase in peak load for uncontrolled charging scenario. However, the results of these studies do not naturally fulfill utilities' needs of quantifying the long-term cost induced by PEV penetration. This is because (i) the existing studies are simulation-based, and thus the conclusions drawn cannot be generalized to other power systems; (ii) TS analysis only shows the electrical response (e.g., voltage, power, etc.), but grid asset depreciation could depend on response in other dimensions (e.g., winding temperature); and most importantly (iii) the load flow resulted from the TS analysis are taken in the form of annual average in the grid asset assessment \cite{abb2018tco}, which makes PEVs' impulsive charging characteristics invisible. In other words, the load spikes caused by PEV charging can be easily averaged off in the assessment and shown harmless, while they could greatly reduce the lifetime of the grid assets in reality. 

To address the above deficiencies, this paper proposes an algorithm to evaluate grid asset depreciation under PEV's penetration. The contributions of the proposed algorithm are twofold:
\begin{itemize}
\item It provides an approach to conveniently assess PEV's impact on grid assets. The PEV charging profiles are pre-processed through Monte Carlo Simulation (MCS), which ensures accounting of random charging patterns, fed into TS analysis and asset lifetime analysis. The outputs are presented through an integrated interface. 
\item Inter-temporal response of grid assets is considered. Compared to existing methods, which assess grid assets based on their average loading, the proposed algorithm considers assets' operating frequency and temperature variation. These factors could lead to significant differences in the assessment, as demonstrated in the numerical cases.
\end{itemize}

The above two engineering advantages are realized under a unified mathematical framework, in which we establish analytical models of two generic classes of grid assets (i.e., continuous and discrete operating assets) and recast their cost functions in the statistical settings of PEV charging. Distinct from simulation-based methods, the proposed method is analytical, and thus greatly reduce the computation resources and data required for accurate assessment. 

The rest of the paper is organized as follows. Section \ref{sec2} introduces the mathematical framework, the analytical models, and the updated cost functions of the grid assets. Section \ref{sec3} demonstrates the effectiveness of the proposed algorithm on a set of power distribution networks in Columbus metropolitan area, Ohio. We further discuss the implications of grid assets' depreciation under different PEV charging settings. Finally, the proposed algorithm and its future applications are concluded in Section \ref{sec4}.

This paper assumes that the power grid operates in the steady-state. The dynamical response of grid assets is defined as the inter-temporal state change. This paper does not address the transient response (i.e., power quality issues) and voltage instability induced by PEV charging \cite{huang2013quasi,mao2018impact}. In the paper, ``grid assets'' and ``power delivery equipment'' are used interchangeably. In addition, although the proposed algorithm can be applied to any power systems, we only examine its effectiveness in simple settings, where mitigation on PEV charging is not applied. An exhaustive examination of PEV's impact on grid assets is out of the scope of this paper. 

\label{sec2}
\subsection{Overview of Proposed Integrated Algorithm}
\begin{figure}[h!]
	\centering
	\includegraphics[scale=0.55]{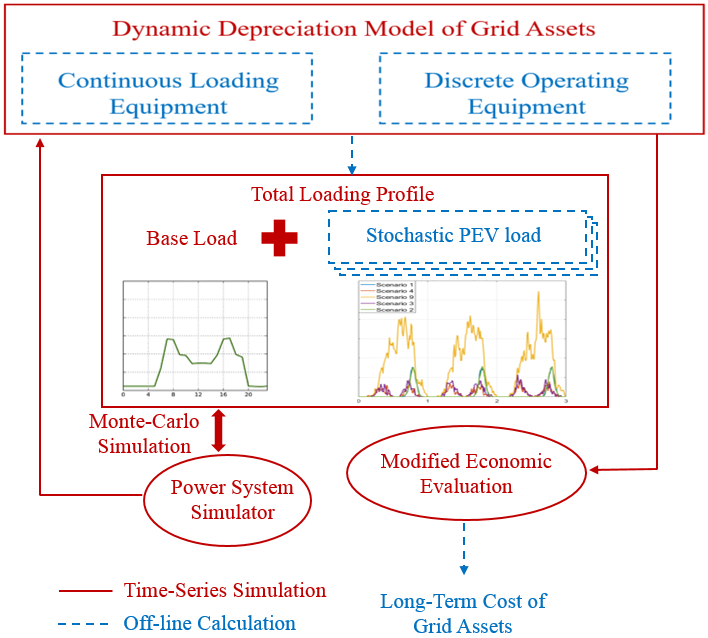}
	\caption{Workflow of Proposed Algorithm}
	\label{flow}
	\vspace{-15pt}
\end{figure}

The proposed integrated algorithm is outlined in Fig.~\ref{flow}. In general, the algorithm combines TS power distribution systems analysis with off-line asset impact assessment. TS analysis is deployed to feed the time-varying grid status to the analytical asset depreciation models. Distinct from existing methods, which approximate actual grid status with annual average values, TS analysis enables accurate evaluation of grid assets' inter-temporal response. MCS is deployed to reflect the stochastic PEV charging patterns in the power flow, which are feed to TS analysis. Based on the Central Limit Theorem, the loading levels output from TS under MCS will provide a more accurate assessment if more charging patterns are available.

\subsection{Total Cost of Ownership Analysis in Utility Practice}
\label{subsec1}
Grid assets can be classified into two categories based on their depreciation procedures: continuous loading equipment and discrete operating equipment. The former's depreciation rate depends on their thermal loading, while the latter's depends on their operating frequency. Examples are transformers, which depreciate faster under heavy loading, and voltage regulators (VR), which exhaust after operating for a certain number of times.

Total Cost of Ownership (TCO) analysis is commonly adopted by utilities to assess the long-term cost, comprised of fixed capital cost and operating depreciation, of power delivery equipment. The TCO of discrete operating equipment is conventionally evaluated independent of loading conditions. For continuous loading equipment, its TCO is exemplified by a transformer and expressed as \eqref{eq:TCO}, with terms expanded in \eqref{eq:A} to \eqref{eq:lof} \cite{rural2016guide}. 
\begin{align}\label{eq:TCO}
\text{TCO} = C_o + CL\cdot A + LL \cdot B,
\end{align}
where $C_o$ is the bid price (capital cost) in dollar of the transformer, the rest of the terms are operating cost in dollar. $CL,LL$ are transformer core loss and load loss provided by manufacturers, $A$ and $B$ are core loss and load loss factor,
\begin{align}
    &A = DC+N\cdot PEC \label{eq:A} \\
    &B = (RF\cdot DC + LoF \cdot PEC) \cdot \hat{P}^2, \label{eq:B}
\end{align}
where $DC$ represents levelized hourly demand cost ([\$/kW-hr]), $N=8760$ is the total hours in a year, $RF$ is the transformer responsibility factor indicating the relationship between transformer peak load and transformer load at a time of system peak, $\hat{P}$ is the normalized peak loading $\hat{s}/s_R$, 
$PEC$ is the present value of energy cost ([\$/kWh]), which depends on the unified transformer insulation life $T_{ins}$, interest rate $i$, and energy cost $EC$ .
\begin{equation}\label{eq:pec}
PEC = EC \cdot \frac{(1+i)^{T_{ins}}-1}{i(1+i)^{T_{ins}}},
\end{equation}
and $LoF$ is transformer loss factor depending on the annual average loading of the transformer $s_{avg}$.
\begin{align}\label{eq:lof}
    LoF = \gamma \frac{s_{avg}}{\hat{s}}+(1-\gamma)(\frac{s_{avg}}{\hat{s}})^2,
\end{align}
where $\gamma$ is the dynamic load factor constant.

In \eqref{eq:TCO}, the last term $LL \cdot B$ models the depreciation induced from transformer loading. From \eqref{eq:B} and \eqref{eq:lof}, it can be seen that average annual loading is used to approximate the time-varying loading. This conventional assessment method can occasionally capture the long-term overloading. However, they are incapable of capturing short-term overloading induced from impulsive PEV loads, because the ``load spikes'' of charging could be easily averaged off.
\subsection{Grid Asset Depreciation Models}
Models of grid assets under the same category take similar forms. Due to limited space, we present the dynamical models of transformers and voltage regulators to represent the continuous loading equipment and discrete operating equipment, respectively, while the proposed algorithm can be generalized to all types of grid assets. These models are adopted to assess the equipment's temporal response in the proposed algorithm. We also derive their corresponding Loss of Life (LoL) models. 
\subsubsection{Continuous Loading Equipment}
The distribution transformer's lifetime depends on the internal winding hot-spot temperature $Q_{HST}$, which is directly related to loading level $s(t)$ at each instant \cite{international2005iec}. The core of this thermal model has the general form in terms of continuous time differential equations:
\begin{eqnarray} \label{eq:qto}
  &\dot{Q}_{TO}(t)=f_1(\mathbb{E}^2[K(t)],Q_{TO}(t))\\ 
    &\ddot{Q}_{H}(t)=f_2(\mathbb{E}^y[K(t)],\dot{Q}_{H}(t))\\ \label{eq:qht}
    &Q_{HST}(t)=Q_{TO}(t)+\tau_H \cdot \dot{Q}_{H}(t), \label{eq:qhst}
\end{eqnarray}
where $Q_{TO}$ is the top-oil temperature, $\mathbb{E}[K(t)]$ is the expectation of load factor $K(t)=s(t)/s_{R}$ (rated) at each instant obtained from distribution power flow analysis embedded with stochastic methodology, $\dot{Q}_{H}$ is the hot-spot temperature dynamic over top-oil, $\tau_H$ is the hot-spot temperature time constant, and $y$ is the winding exponent power. The compact form of the dynamical system model of \eqref{eq:qto}-\eqref{eq:qhst} can be written  as a stochastic function of continuous loading level.
\begin{align}
    &\dot{Q}_{X}=f(Q_X,s(t)|\mu, \sigma)\\
    &Q_{HST}=a^T \cdot Q_X,
\end{align}
where $Q_X=[Q_{TO}~\dot{Q}_{H}]$ and $a=[1 ~\tau_H]^T$.

Then, the actual loss of life $L_T$ for transformer during any time span $[t_1,t_2]$ is derived as \eqref{eq:lt}. The transformer's expected lifetime $T_x$ can be found by solving $L_x(0,T_x)=1$. 
\begin{align}\label{eq:lt}
L_x (t_1,t_2)=\frac{1}{T_{ins}} \int \nolimits_{t_1}^{t_2} F_{AA}(t)dt,
\end{align}
where $T_{ins}$ is the normal insulation life of the transformer and $F_{AA}$ is the accelerated aging factor defined in \eqref{eq:faa} \cite{board1995ieee}. When $F_{AA}(t)>1$, the lifetime of the transformer is shortened at instant $t$.
\begin{align}
    &F_{AA}(Q_{HST})=exp(\alpha-\frac{\beta}{Q_{HST}(t)+\Omega}), \label{eq:faa}
\end{align}
where $\alpha,~\beta$ and $\Omega$ are design constants of the transformer.
\subsubsection{Discrete Operation Equipment}
Voltage regulators (VR) are essentially a type of tap changing transformer. In the distribution level of power grid, VR are used to regulate voltage deviation from predetermined values. Impulse loads, like PEV, tend to cause fast time-varying and salient voltage deviation, which may result in more frequent operation of VR. VR's lifetimes are determined by their mechanical durability and specified as the total number of effective tap operations. The operation policy of VR can be expressed as \eqref{eq:torus}.
\begin{equation}
h(n)= 
 \begin{cases}
  (V(n)-V_{R})\cdot \frac{1}{\kappa}, \text{ if } V(n)\in[h_{min},\underline{\epsilon}]\cup[\bar{\epsilon},h_{max}]\\
    h_{max}, \qquad \text{if } h(n-1)+\Delta h(n)\geq h_{max}\\
    h_{min}, \qquad \text{if } h(n-1)-\Delta h(n)\leq h_{min}\\
  \end{cases}, \label{eq:torus}
\end{equation}
where $h(n)$ is the VR tap position at the $n^{th}$ sampled instant after each operating cycle, $V(n)$ is the discrete voltage level calculated from power flow, $V_R$ is the regulated voltage, $\kappa$ is the VR step-size, $[\underline{\epsilon},\bar{\epsilon}]$ is VR's dead-band, and $h_{max}$, $h_{min}$ are maximum and minimum tap position.

By observing such change of tap positions triggered by voltage variation, the LoL of VR during any time span $[n_1,n_2]$ can be obtained in \eqref{eq:lvr}, and the VR's lifetime $T_v$ can be founded by solving $L_{v}(0,T_v)=1$.
\begin{align}\label{eq:lvr}
L_{v}(n_1,n_2)=\frac{1}{N_{op}}\sum_{n_1}^{n_2} |h(n)-h(n-1)|,
\end{align}
where $N_{op}$ is the VR's empirical maximum number of tap operations.
\subsubsection{Re-established TCO Evaluation}
\label{subsubsec_tco}
The outputs of TS analysis enable us to accurately assess the LoL of power delivery equipment in the grid with PEV loads during any time span of interest. In this section, we re-establish the TCO formulation for grid asset long-term cost assessment. For VR, the TCO can be simply expressed as
\begin{equation}\label{eq:mtcovr}
\text{TCO}(n_1,n_2)=L_{V}(n_1,n_2)\cdot C_o,
\end{equation}
where $L_{V}(n_1,n_2)$ is specified in \eqref{eq:lvr} and $C_0$ is the VR's capital cost. 
For transformers, the TCO can be formulated as

\begin{align}\label{eq:mtcot}
\text{TCO} (t_1,t_2) = &L_x (t_1,t_2)\cdot C_o \\ \nonumber
&+ CL\cdot A(t_1,t_2) + LL \cdot B(s,t_1,t_2),
\end{align}
where $L_x(t_1,t_2)$ is specified in \eqref{eq:lt} and other parameters are specified in Section \ref{subsec1}. $PEC$ in \eqref{eq:pec} is modified to reflect the future cost in $[t_1,t_2]$ to the present day value as 
\begin{equation}
PEC = \frac{EC}{i}[\frac{1}{(1+i)^{t_1}}-\frac{1}{(1+i)^{t_2}}],
\end{equation}
and the parameter $LoF$ in \eqref{eq:lof} is modified to capture time-varying loading level under stochastic PEV charging patterns as
\begin{align}
    LoF(s,t) = \gamma \frac{\mathbb{E}[s(t)]}{\hat{s}}+(1-\gamma)(\frac{\mathbb{E}[s(t)]}{\hat{s}})^2 .
\end{align}

In both \eqref{eq:mtcovr} and \eqref{eq:mtcot}, the first term reflects the capital cost of the equipment due to the accelerated depreciation resulted from extra stress of PEV loads, while the other terms in \eqref{eq:mtcot} reflects the operating cost induced from stochastic TS load profiles. Therefore, the re-established TCO evaluation, with TS analysis and the two analytical models, can accurately capture any overloading form.

\section{Case Study}
\label{sec3}
To demonstrate the validity of the proposed algorithm, case studies of real-world distribution grids are carried out in this section. Simulation results of grid asset depreciation state and long-term cost evaluation are presented.
\subsection{Overview of System Topology and Simulation Setup}
\begin{figure}[h!]
\begin{subfigure}{0.325\textwidth}
\includegraphics[width=1\linewidth, height=10cm]{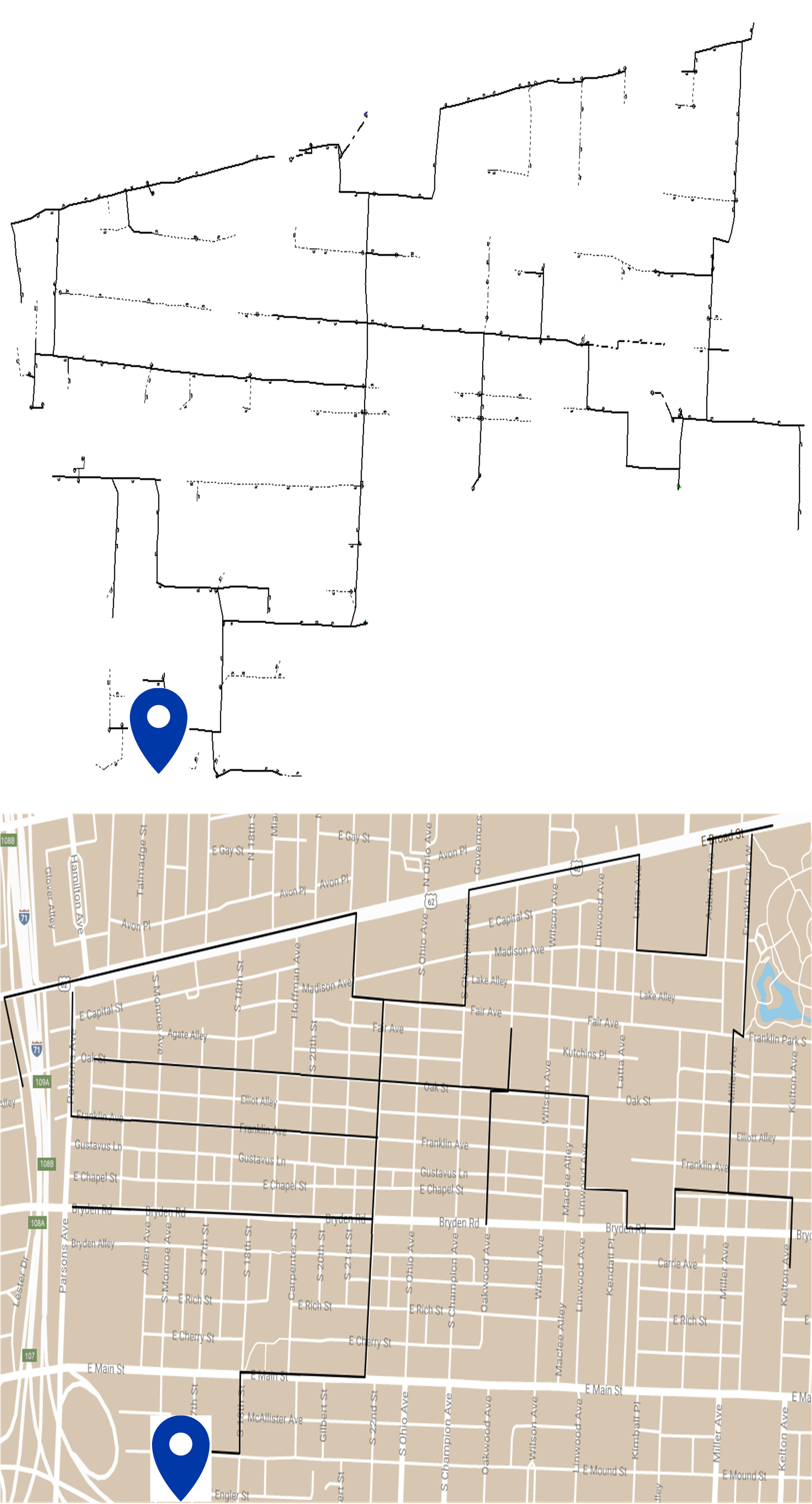} 
\caption{Urban Area}
\label{fig:subim2}
\end{subfigure}
\begin{subfigure}{0.325\textwidth}
\includegraphics[width=1\linewidth, height=10cm]{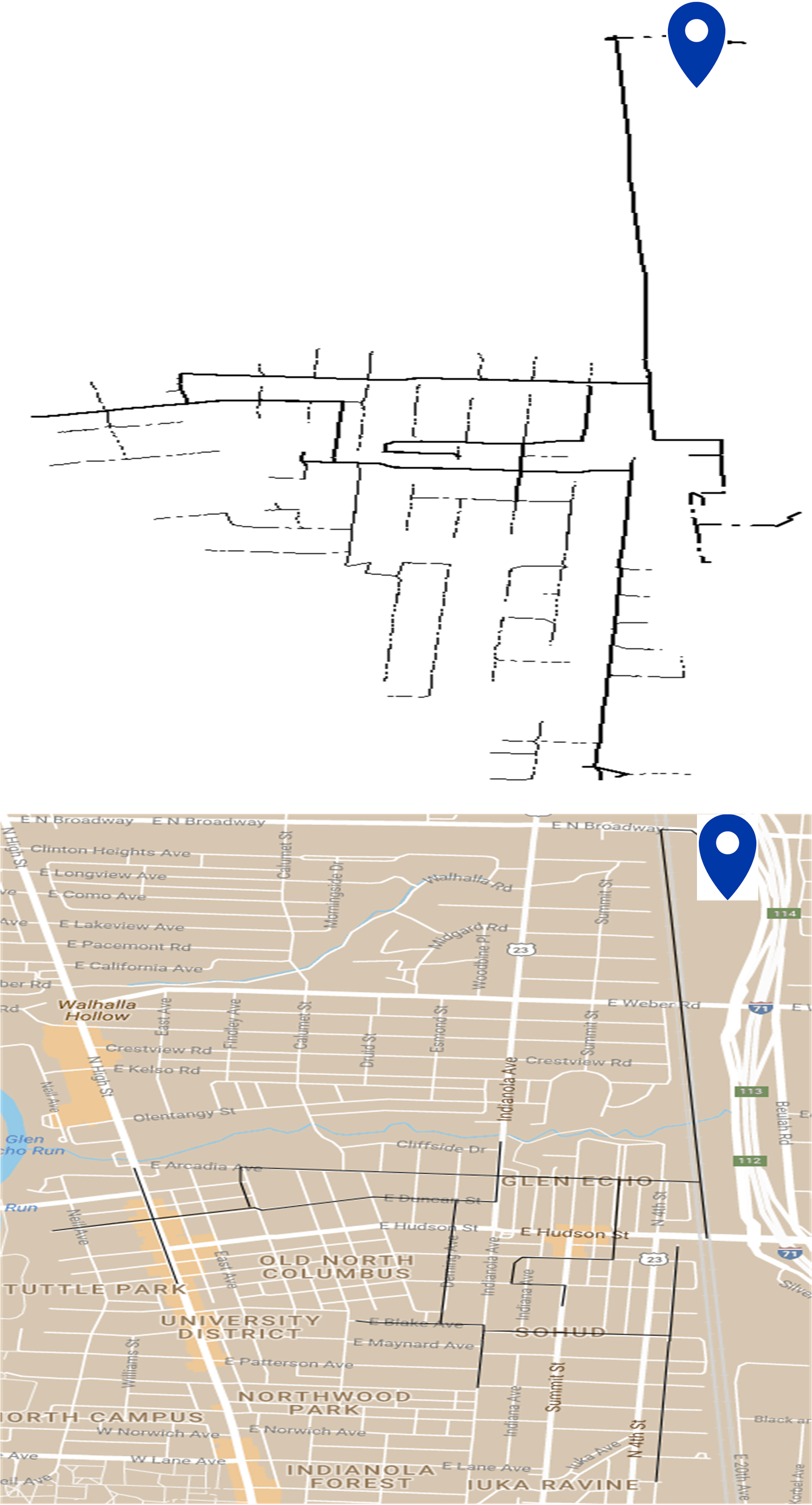} 
\caption{Suburban Area}
\label{fig:subim1}
\end{subfigure}
\begin{subfigure}{0.325\textwidth}
\includegraphics[width=1\linewidth, height=10cm]{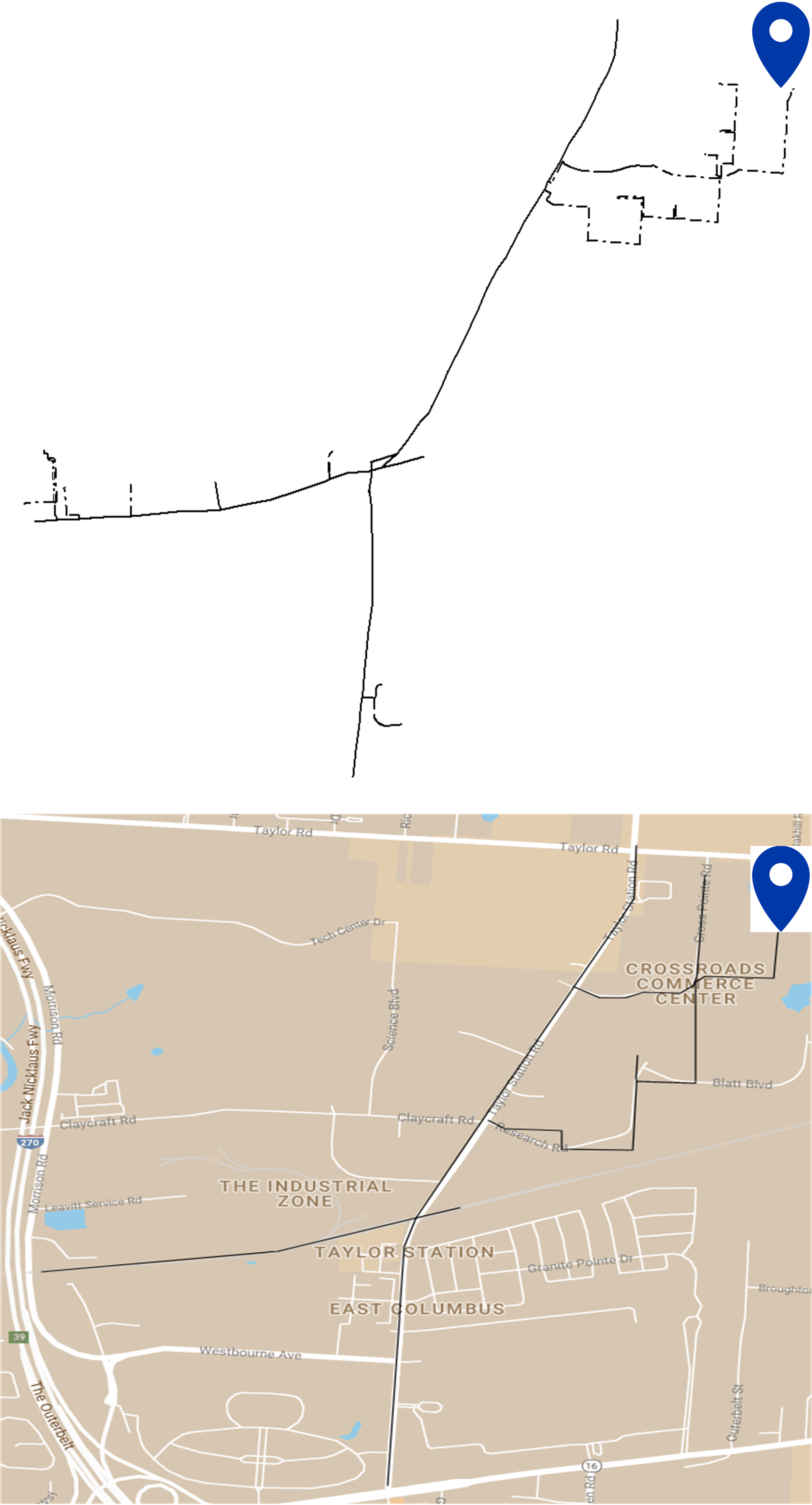}
\caption{Rural Area}
\label{fig:subim3}
\end{subfigure}
\caption{Topology of Power Distribution Networks}
\label{topology}
\end{figure}

The integrated methodology outlined in Section \ref{sec2} has been applied to three large-scale power distribution systems in Columbus metropolitan area, Ohio. These three areas can be demographically categorized as in Table \ref{t_area}.
\begin{table}[h!]
\begin{center}
\caption {Demographic Categorization of Distribution Networks} \label{t_area} 
    \begin{tabular}{ | c | c | c | }
    \hline
    Community & Electric Service Area (km$^2$) & Connected Capacity (kVA)\\ \hline
    Urban & 2.820 & 16793\\ \hline
    Suburban & 5.568 & 11661\\ \hline
    Rural & 6.786 & 7707\\ \hline
    \end{tabular}
    \vspace{-20pt}
\end{center}
\end{table}

The above areas have a comparable amount of base demand, i.e., connected capacity. The electric circuit data obtained from American Electric Power (AEP) is originally formatted in CYME, a commercial-grade power system simulation software widely used by electric utilities. Due to the customized simulation setup and the need for flexible PEV load integration, all data has been first converted to the format in OpenDSS \cite{modelopendss}, an open-source power distribution system simulator. The entire algorithm hereafter is demonstrated with MATLAB and OpenDSS.

The topology of the electric circuit from CYME and corresponding network atlas from Google Maps are shown in Fig. \ref{topology}. The main feeder of power distribution circuit has been sketched in the map by black solid line and the locations of the substation for each area have been labeled by blue marker for illustrative purpose. The urban circuit has the longest main feeder and the highest density of sub-feeders throughout the network, whereas the suburban circuit has a relatively sparse distribution of sub-feeders, followed by the rural circuit which has a simple tree topology and the lowest sub-feeder density.

\begin{figure}[h!]
\vspace{-5pt}
\begin{subfigure}{0.325\textwidth}
\includegraphics[width=1\linewidth, height=5cm]{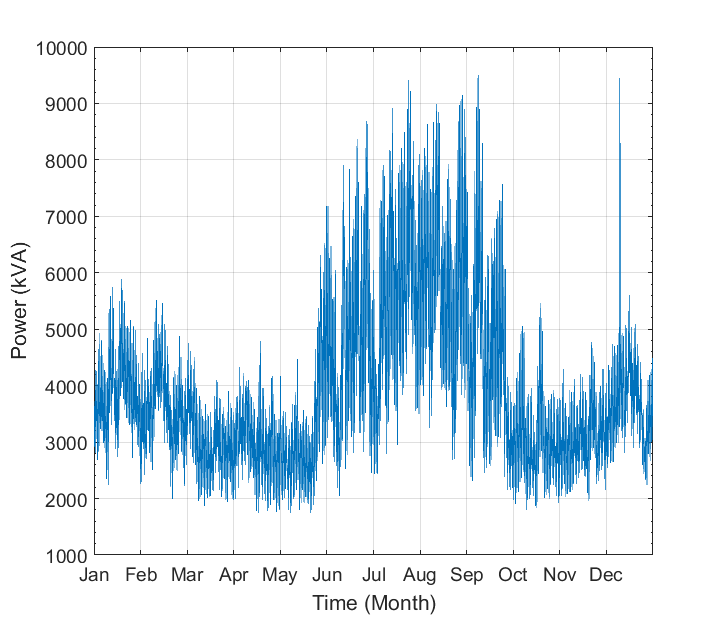} 
\caption{Urban Area}
\label{noevsub1}
\end{subfigure}
\begin{subfigure}{0.325\textwidth}
\includegraphics[width=1\linewidth, height=5cm]{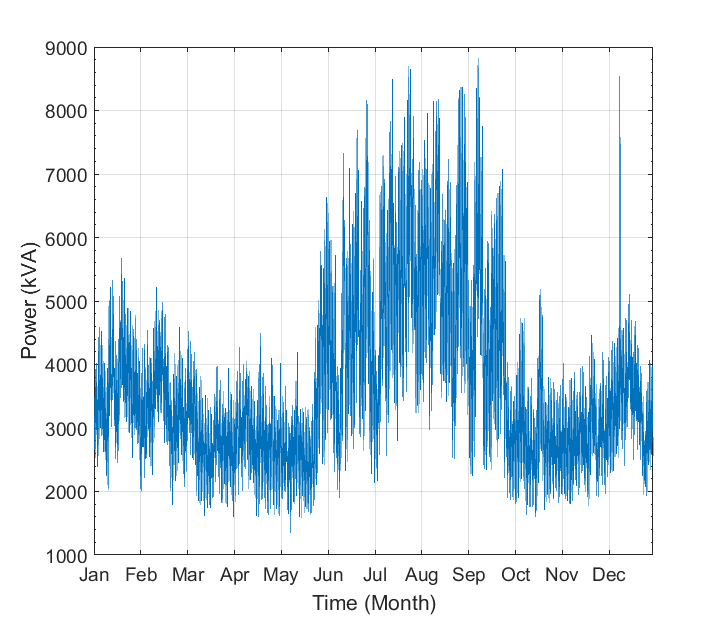} 
\caption{Suburban Area}
\label{noevsub2}
\end{subfigure}
\begin{subfigure}{0.325\textwidth}
\includegraphics[width=1\linewidth, height=5cm]{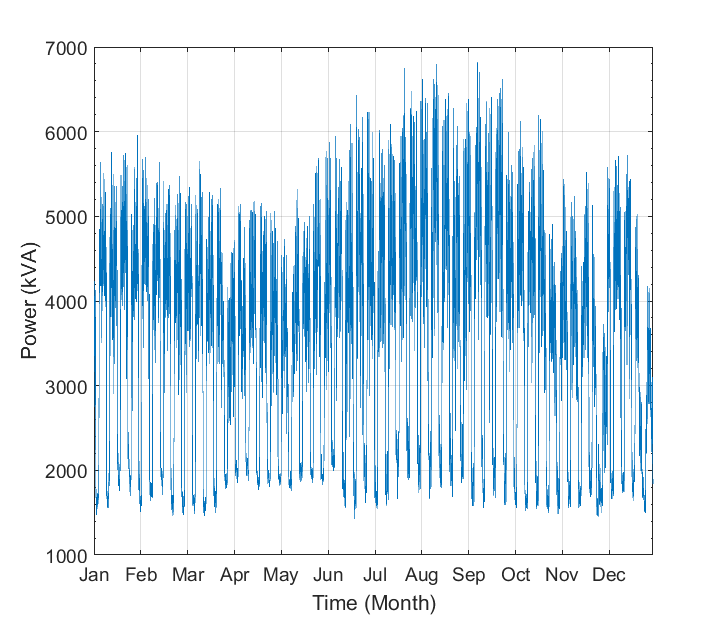}
\caption{Rural Area}
\label{noevsub3}
\end{subfigure}
\caption{Yearly Profile of Base Load}
\label{noev}
\vspace{-15pt}
\end{figure}

In terms of load condition, the ``base load'' shown in Fig. \ref{noev} serves as the benchmark in our case study. It is recorded at the substation of each area in a 15-minutes resolution for one year and it is assumed that no PEVs are connected in this benchmark case. In addition to circuit configuration, these three areas also differ in loading demographics. The urban and suburban circuits are mainly comprised of residential and commercial load type, whereas the industrial load type dominates the rural circuit. As shown in Fig. \ref{noevsub3}, the envelope of the load profile in the rule area is stretched wider because some industrial loads usually are constantly running at their full capacity during the work time while completely off during the night, weekend, and holidays. 

\subsection{Simulated PEV Charging Scenarios}
\label{subsec3.2}
The total load at any location $h$ in the network is the summation of the base load $P_h^b$ and aggregated PEV load $P_h^{PEV}$, i.e., $P_h(t)=P_h^b (t)+\sum P_h^{PEV}(t)$. There are multiple factors that collectively affect individual PEV's daily charging profile $P^{PEV}(t)$. In this case study, the following three typical aspects are considered: (i) charging level; (ii) battery capacity; and (iii) vehicle type.

For each aforementioned aspect, two specifications are assumed. The PEV can utilize ``slow-charging'' level $P=19.2~kW$, which is commonly used in the residential household as expedited home charging level, or ``fast-charging'' level $P=120~kW$, which is a widely used DC public charging level exemplified by Tesla supercharger \cite{yilmaz2013reviewbattery,tan2016comprehensive}. The battery capacity are assumed to be ``short-range'' $C=40 ~kWh$ or ``long-range'' $C=60 ~kWh$.\footnote{The 2018 Nissan LEAF is equipped with 40 kWh battery pack and the 2018 Chevrolet Bolt EV is equipped with 60 kWh battery pack.} As for the vehicle type, we assume that the PEV is either used as commuter or as ride-service (e.g., Uber, shuttle, cab, etc.). 

The charging level and vehicle type collaboratively determine the stochastic PEV charging behavior, which is modeled by two random variables: the charging start time $t_{s}$ and the charging period $\Delta t$. The latter is an explicit function of initial State of Charge (SoC) of battery at the beginning of each charging action, given battery capacity $C$ and charging level $P$, i.e., $\Delta t=C\cdot(1-SoC)\big{/}P$.

In terms of vehicle types, for the commuter who utilizes slow-charging, we assume that charging occurs right after getting back home from work, and the charging period $\Delta t$ is determined by the mileage driven for commute each day. According to National Household Travel Survey \cite{santos2011summary}, the individual commuter's departure/arrival time and daily driving mileage are assumed to follow Normal distribution. For the commuter who utilizes fast-charging, we assume that charging occurs either en route to work or on the way home. Whether or not the PEV charge en route is determined by a range anxiety threshold $\tau=30\%$, as opposed to slow-charging commuter case where every PEV charges at home every night. The ride-service type of PEVs will be driving for daily service from 7 am to 9 pm and also charge en route whenever the SOC falls below the threshold $\tau$. The average speed for different time period in a day is used to formulate their multiple charging need \cite{gao2018visualizing}. Noted that the ride-service type of PEV is only considered to utilize fast-charging level and be equipped with long-range battery due to the inherent requirement of vehicle usage. From an aggregation point of view, it is assumed that each area studied has either 500 or 1000 PEV fleets in order to observe progressive impact. The number of fleets simulated in this case study is consistent with the penetration goal set by the U.S. that every household owns a PEV in the future \cite{singer2016consumer,block2017prediction}. For some states such as California, the PEV penetration goal in coming decades has been even more aggressive, as almost 2 fleets per household \cite{searle2016leading,lutsey2018california}. Accordingly, the particular case of 500 and 1000 PEVs fall in the reasonable median of the PEV density. All simulated charging scenarios determined by aforementioned factors are summarized and indexed in Table \ref{t2}.
\begin{table}[h!]
\begin{center}
\caption {Summary of Simulated PEV Charging Scenarios} \label{t2} 
    \begin{tabular}{ | c | c | c | c | c |}
    \hline
    Vehicle Type & No. of Fleets & Charging Level & Battery Capacity & Scenario Index \\ \hline
    \multirow{8}{*}{Commuter} & \multirow{4}{*}{500} & \multirow{2}{*}{Slow-charging} & Short-range &1\\ \cline{4-5}
    &  &  & Long-range &2\\ \cline{3-5}
    &  & \multirow{2}{*}{Fast-charging } & Short-range &3\\ \cline{4-5}
    &  &  & Long-range &4\\ \cline{2-5}
    & \multirow{4}{*}{1000} & \multirow{2}{*}{Slow-charging } & Short-range  &5\\ \cline{4-5}
    &  &  & Long-range &6\\ \cline{3-5}
    &  & \multirow{2}{*}{Fast-charging } & Short-range  &7\\ \cline{4-5}
    &  &  & Long-range  &8\\ \hline
    \multirow{2}{*}{Ride-service} & 500 & \multirow{2}{*}{Fast-charging } & \multirow{2}{*}{Long-range } &9\\ \cline{2-2} \cline{5-5}
    & 1000 &  & &10 \\ \hline
    \end{tabular}
\end{center}
\end{table}

\begin{figure}[h!]
	\centering
	\includegraphics[scale=0.6]{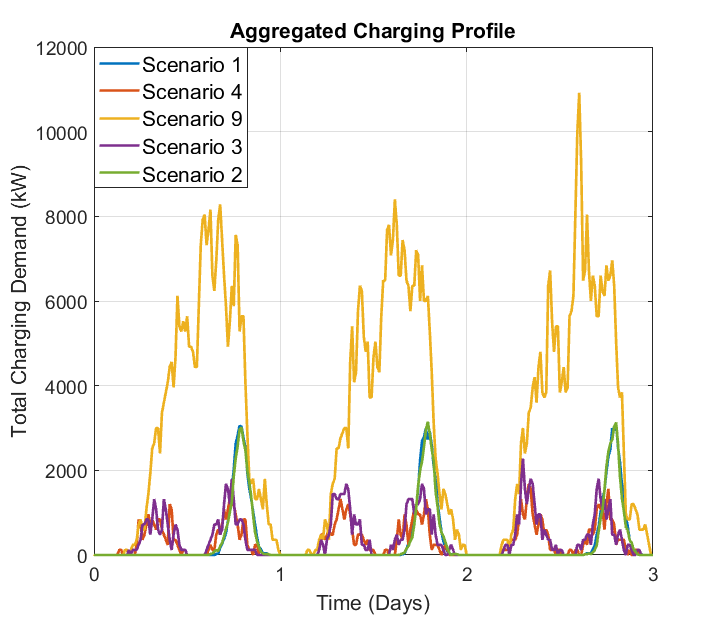}
	\caption{Aggregated PEV Charging Profile of Random 3-day Period}
	\label{aggregate}
\end{figure}

The aggregated PEV charging profile of 500 fleets in a randomly selected 3-day period for all charging scenarios (viz. Table \ref{t2}) is shown in Fig. \ref{aggregate}. It can be seen that even though the slow-charging commuter has lower charging power individually, it's much easier for them to have coincidental charging than fast-charging commuter scenario due to the concentration of home charging events. Utilizing fast-charging level, the PEV only needs to recharge every 3-4 days and has a shorter period needed for each charging action. Moreover, the en route fast-charging actions have been split equally into departure en route and arrival en route charging, i.e., the aggregated daily charging of fast-charging commuter has two spikes as compared to the single higher spike of slow-charging commuter. On the other hand, the fast-charging ride-service scenario reveals the most significant loading condition among all scenarios. Individual ride-service PEV will be charging en route multiple times (1$\sim$4) during their service hours every day. The battery capacity imposes less influence on aggregated charging profile than the other two factors.

To truly reflect stochastic PEV charging patterns, Monte-Carlo Simulation (MCS) is implemented in TS analysis. The power flows in the grid are simulated in multiple iterations under total TS load profiles and are input into the analytic models of grid asset response simultaneously. The law of large numbers indicates that as the sample size gets large enough, the expected value of model outputs can be approximated by taking the sample mean of the MCS output results. For example, the $\mathbb{E}[K(t)]$ in (\ref{eq:qto}) and (\ref{eq:qht}) is the \emph{expectation} of load factor $K(t)$ at each instant obtained from averaging the power flow result over MCS iterations. In this case study, every charging scenario has been iterated 100 times in MCS, with 500 or 1000 fleets of PEV yearly charging profiles randomly allocated in the area during each iteration.

\subsection{Grid Asset Depreciation Analysis}
\label{subsec_grid}
\subsubsection{Transformer Depreciation induced by PEV Charging}
This section presents the lifetime depreciation evaluation of substation transformer induced by PEV charging. The thermal parameters related to transformer LoL estimation are collectively selected from \cite{international2005iec,board1995ieee,sen2011transformer}. Noted that the substation transformers are assumed to have the same rating $s_R=10$ MVA and thermal parameters in all three areas due to the lack of field measurement and \textit{ceteris paribus} need. Moreover, the normal insulation life of substation transformer has been selected as 25\% retained tensile strength $T_{ins}=15.41$ yr ($135,000$ hr) \cite{international2005iec}.

\begin{figure}[h!]
\begin{subfigure}{0.325\textwidth}
\includegraphics[width=1\linewidth, height=5.5cm]{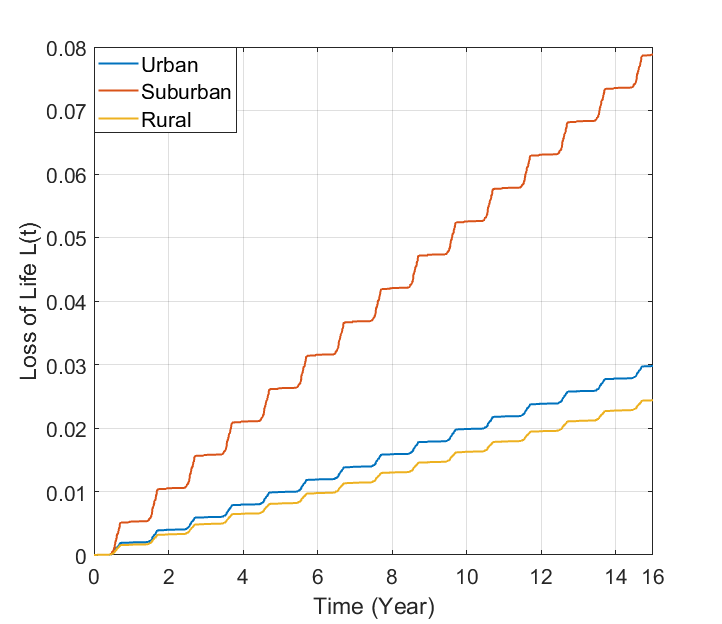} 
\caption{Benchmark}
\label{sub1}
\end{subfigure}
\begin{subfigure}{0.325\textwidth}
\includegraphics[width=1\linewidth, height=5.5cm]{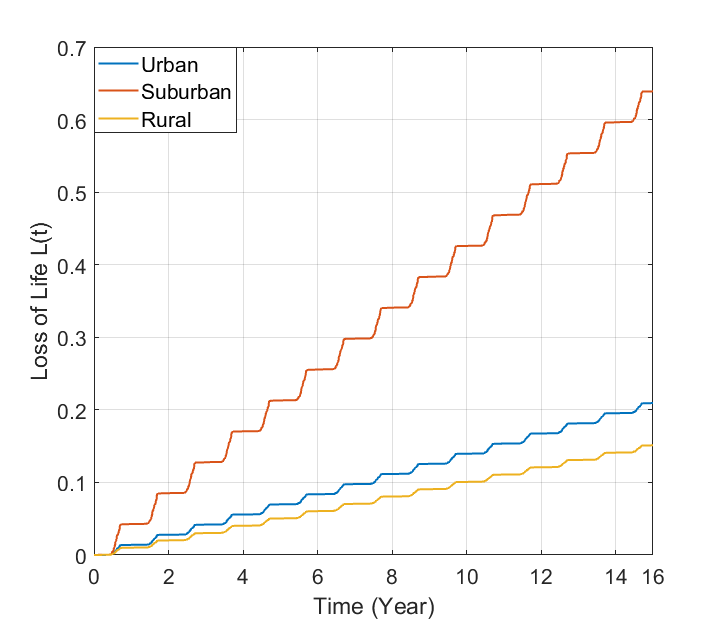} 
\caption{Scenario 1}
\label{sub2}
\end{subfigure}
\begin{subfigure}{0.325\textwidth}
\includegraphics[width=1\linewidth, height=5.5cm]{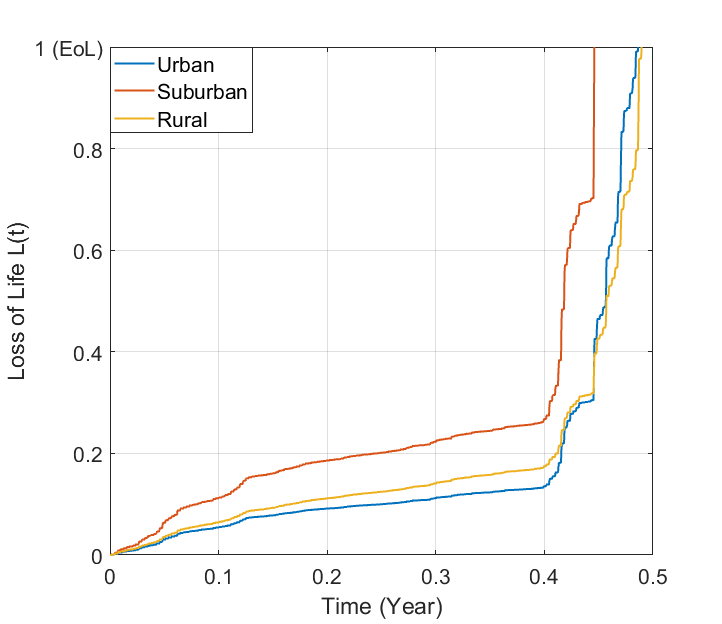} 
\caption{Scenario 9}
\label{sub3}
\end{subfigure}
\caption{Accumulated Transformer Loss of Life}
\label{LoL_combined}
\end{figure}

The transformer's accumulated LoL based on thermal model \eqref{eq:qto} - \eqref{eq:lt} are exemplified by Fig. \ref{LoL_combined}, demonstrating two specific charging scenarios and benchmark case for all areas. The abrupt increase of degradation after 0.4 year in each one-year period is mainly attributed to the shape of base load as shown in Fig. \ref{noev}. All three areas' base load have the similar pattern that the envelope of load curve starts to stretch upward in mid-May. As the PEV load profile simulated in the paper has no seasonal fluctuation, the total load profile pattern will be consistent with base load. Hence, the degradation will start to speed up as it is highly related to the loading level.

When the same charging scenario applied to different demographic areas, it can be seen that the transformer in suburban area is most prone to induce depreciation, followed by urban and rural area, thus revealing the grid's topological impact under the same charging scenario. On the other hand, when different charging scenarios applied to the same area, it can be observed that the ride-service type of PEV will induce more drastic burden to the asset than commuter type (cf. Fig. \ref{sub2} \& Fig. \ref{sub3}). The utility must consider the upgrade of transformer to a higher rating. Otherwise, according to the asset depreciation model, the current 10 MVA transformer will endure an extremely high overloading burden that makes the transformer reach End of Life (EoL) within a year.

\begin{figure}[h!]
	\centering
	\includegraphics[scale=0.35]{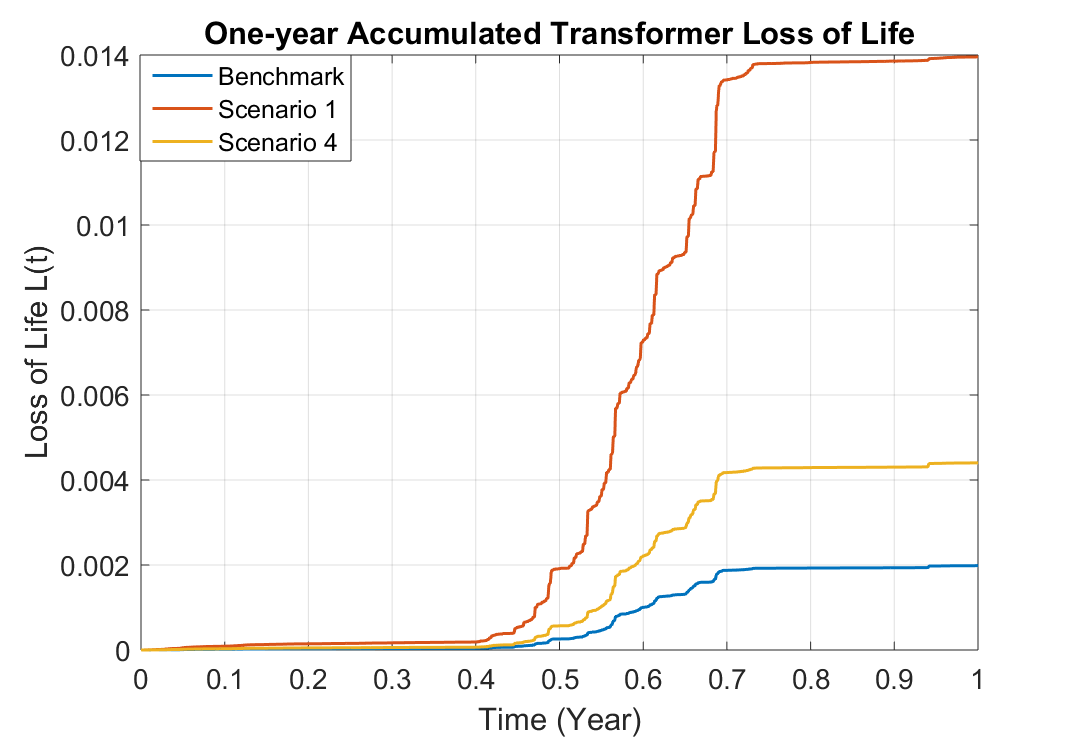}
	\caption{Zoom-in Look of One-year LoL in Urban Area}
	\label{yearlycompare}
\end{figure}

Fig. \ref{yearlycompare} shows a zoom-in look of LoL pattern in a one-year period, comparing benchmark with scenario 1 and 4. It can be seen that the rate of LoL, i.e., the stiffness of LoL curve, is increased under impulsive PEV charging load, thus the lifetime of transformer will be greatly shortened. Moreover, the scenario 1 has a more detrimental effect to transformer than scenario 4 does due to the aforementioned concentration of home charging events.

The transformer depreciation evaluation of all simulated charging scenarios is summarized in Table \ref{t4}. The Yearly LoL indicates the percent loss of life with respect to normal insulation life $T_{ins}$ per year, thus the estimated lifetime can be obtained as $100/(\text{Yearly LoL})$. If this value is longer than 15.41 yr, then the corresponding charging scenario is considered to have no noticeable impact on transformer. On the other hand, there are several charging scenarios for each area that will greatly reduce transformer's lifetime. For certain drastic case such as ride-service type of PEV charging, the estimated lifetime can even be shortened to $\varepsilon<0.5$ yr, which shows an urgent need for the upgrade of critical equipment. All charging scenarios that impose such salient impact on transformer lifetime and thus considered to be unacceptable for utilities are marked in shade in the Table \ref{t4}. 

Noted that the LoL is very sensitive to the transformer rating and certain thermal parameters, thus the estimated lifetime under each charging scenario only falls in a ballpark range. Therefore, besides referring to the estimated lifetime as an absolute reference, another dimension of asset state assessment is to compare the charging scenario's relative LoL with each other, due to the consistent pattern of LoL for all simulated charging scenarios.

\begin{table}[h!]
\begin{center}
\caption {Summary of Estimated Transformer Lifetime} \label{t4} 
    \begin{tabular}{ | c | c | c | c |}
    \hline
    Area & Charging Scenario Index & Yearly LoL (\%) & Lifetime (yr) \\ \hline
    \multirow{11}{*}{Urban} & Benchmark & 0.19 & 15.41 \\ \cline{2-4}
    & 1 & 1.36  & 15.41 \\ \cline{2-4}
    & 2 & 1.18  & 15.41 \\ \cline{2-4}
    & 3 & 0.53 & 15.41 \\ \cline{2-4}
    & 4 & 0.43  & 15.41 \\ \cline{2-4}
    & 5 & 27.64 & \cellcolor{Gray} 3.62 \\ \cline{2-4}
    & 6 & 19.69 & \cellcolor{Gray} 5.08 \\ \cline{2-4}
    & 7 & 1.63 E+03 & \cellcolor{Gray} $\varepsilon$ \\ \cline{2-4}
    & 8 & 1.01  & 15.41 \\ \cline{2-4}
    & 9 & 1.05 E+03  & \cellcolor{Gray} $\varepsilon$ \\ \cline{2-4}
    & 10 & 1.82 E+06  & \cellcolor{Gray} $\varepsilon$ \\ \hline
    \multirow{11}{*}{Suburban} & Benchmark & 0.51 & 15.41 \\ \cline{2-4}
    & 1 & 4.15  & 15.41 \\ \cline{2-4}
    & 2 & 3.58 & 15.41 \\ \cline{2-4}
    & 3 & 1.52 & 15.41 \\ \cline{2-4}
    & 4 & 1.31 & 15.41 \\ \cline{2-4}
    & 5 & 93.81 & \cellcolor{Gray} 1.07 \\ \cline{2-4}
    & 6 & 66.26 & \cellcolor{Gray} 1.51 \\ \cline{2-4}
    & 7 & 24.73 & \cellcolor{Gray} 4.04 \\ \cline{2-4}
    & 8 & 2.98 & 15.41 \\ \cline{2-4}
    & 9 & 3.22 E+03 & \cellcolor{Gray} $\varepsilon$ \\ \cline{2-4}
    & 10 & 6.56 E+06 & \cellcolor{Gray} $\varepsilon$ \\ \hline
    \multirow{11}{*}{Rural} & Benchmark & 0.16 & 15.41 \\ \cline{2-4}
    & 1 & 0.98 & 15.41 \\ \cline{2-4}
    & 2 & 0.86 & 15.41 \\ \cline{2-4}
    & 3 & 0.40 & 15.41 \\ \cline{2-4}
    & 4 & 0.30 & 15.41 \\ \cline{2-4}
    & 5 & 19.40 & \cellcolor{Gray} 5.15 \\ \cline{2-4}
    & 6 & 13.80 & \cellcolor{Gray} 7.25 \\ \cline{2-4}
    & 7 & 5.22 & 15.41 \\ \cline{2-4}
    & 8 & 0.74 & 15.41 \\ \cline{2-4}
    & 9 & 0.66 E+03 & \cellcolor{Gray} $\varepsilon$ \\ \cline{2-4}
    & 10 & 1.63 E+06 & \cellcolor{Gray} $\varepsilon$\\ \hline
    \end{tabular}
\end{center}
\end{table}

\subsubsection{Voltage Regulator Depreciation induced by PEV Charging}
\begin{figure}[h!]
\begin{subfigure}{0.325\textwidth}
\includegraphics[width=1\linewidth, height=5.5cm]{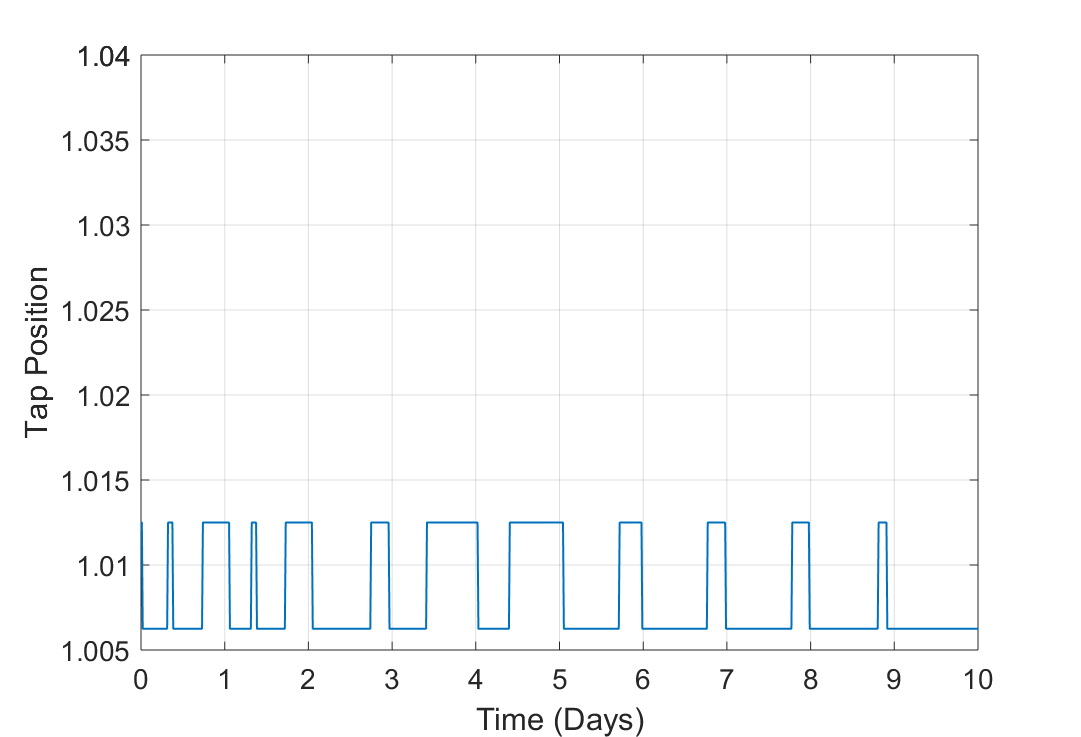} 
\caption{Benchmark}
\label{vrdailybench}
\end{subfigure}
\begin{subfigure}{0.325\textwidth}
	\includegraphics[width=1\linewidth, height=5.5cm]{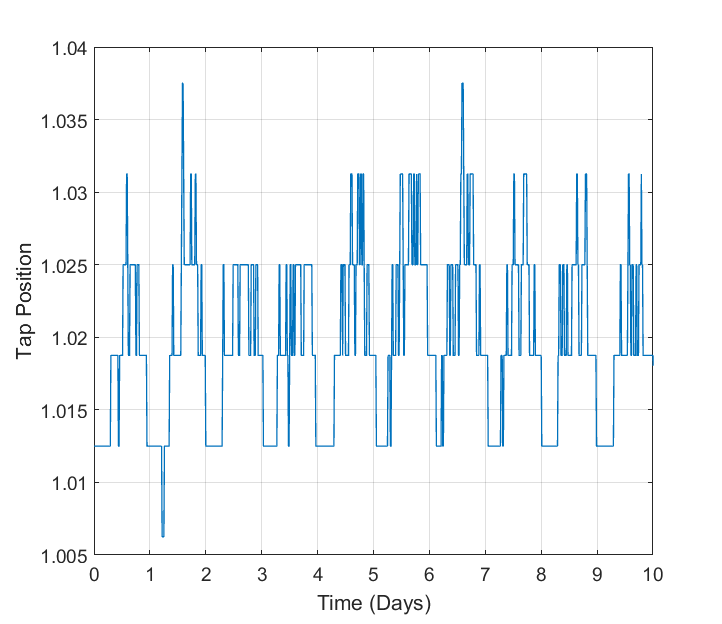}
	\caption{Scenario 9}
	\label{vrdaily500}
\end{subfigure}
\begin{subfigure}{0.325\textwidth}
	\centering
	\includegraphics[width=1\linewidth, height=5.5cm]{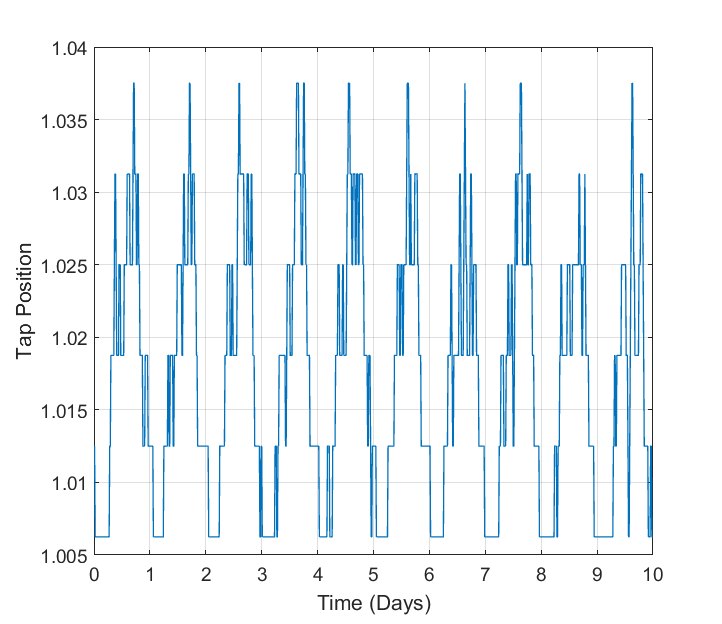}
	\caption{Scenario 10}
	\label{vrdaily1000}
	\end{subfigure}
\caption{VR Tap Operation of Random 10-day Period in Urban Area}
\label{vrdaily}
\end{figure}

This section presents the asset state evaluation of voltage regulator (VR) induced by PEV charging. We assume that there are two three-phase VRs installed at midway through the feeder in each area. The step-size of VR tap is selected as $\kappa=0.0065$. The yearly simulation monitored their tap operations. Fig. \ref{vrdaily} shows the tap operation in a randomly selected 10-day period for urban area, comparing the VR operating frequency in terms of number of PEV fleets with the benchmark. The temporal response of VR is strongly correlated with PEV daily charging activities. Moreover, the increasing number of PEV causes greater voltage deviation and more salient TS load profile, thus induce more frequent VR tap operations.

\begin{figure}[ht]
\begin{subfigure}{0.325\textwidth}
\includegraphics[width=1\linewidth, height=5.5cm]{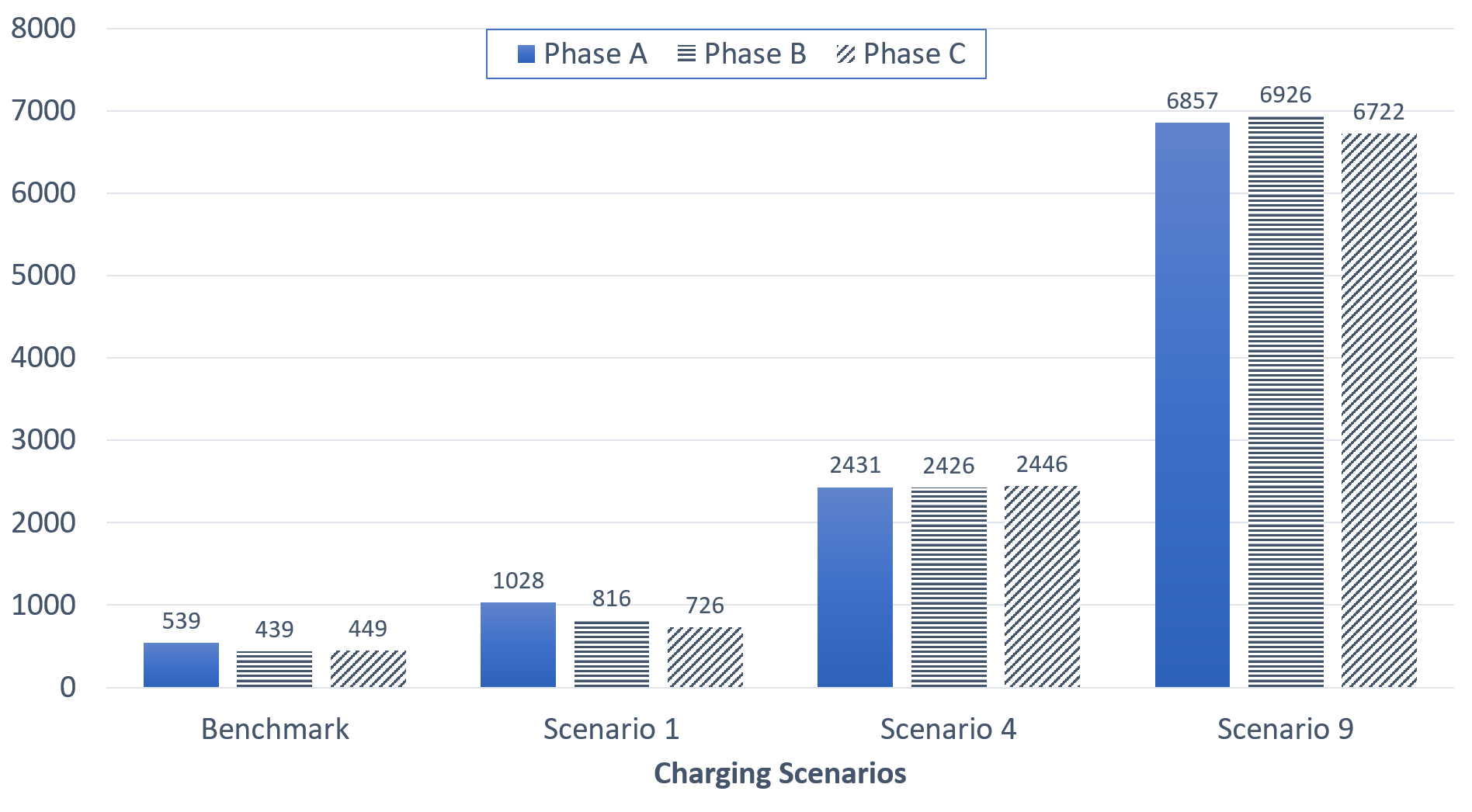} 
\caption{Urban Area}
\label{fig:subim2}
\end{subfigure}
\begin{subfigure}{0.325\textwidth}
\includegraphics[width=1\linewidth, height=5.5cm]{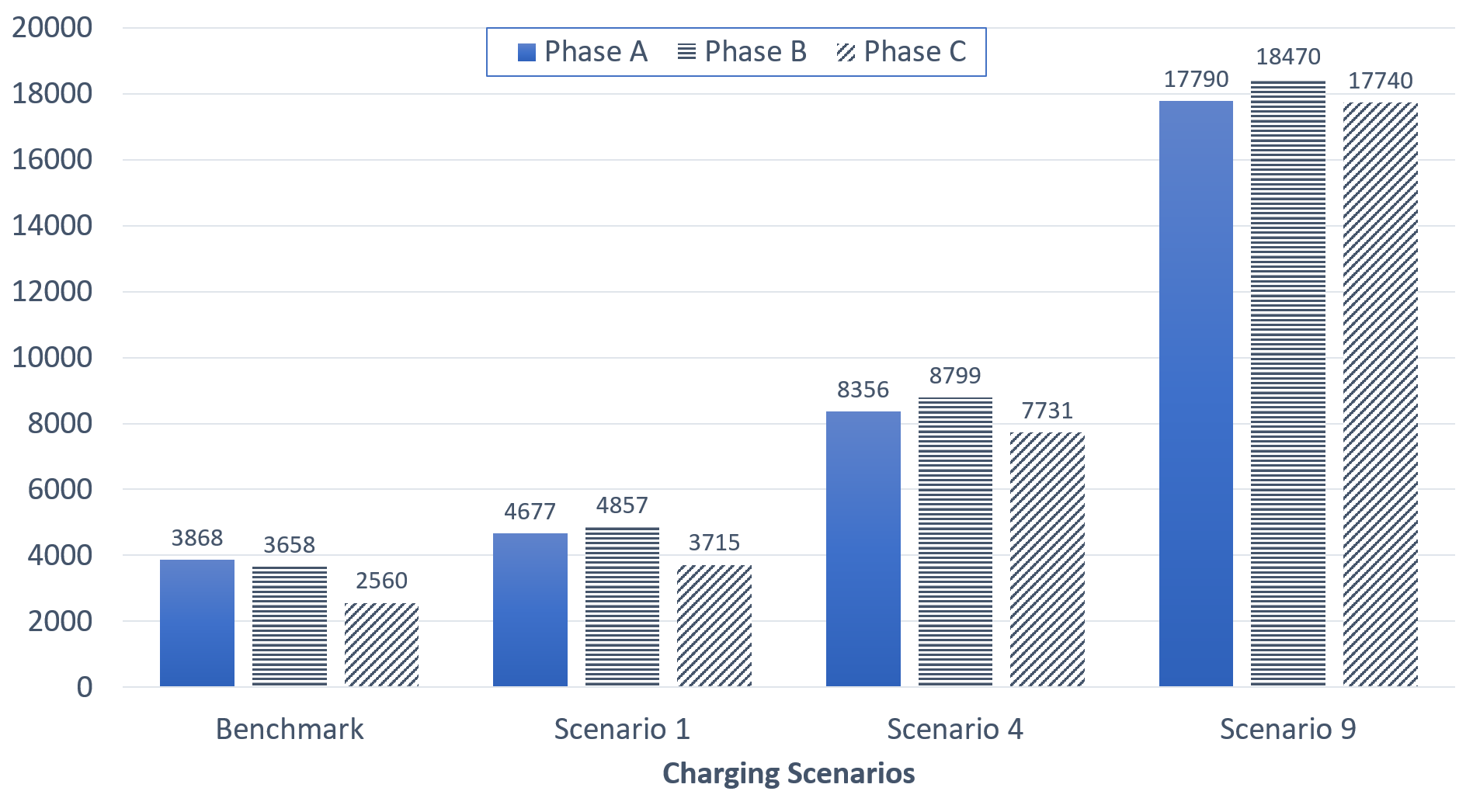} 
\caption{Suburban Area}
\label{fig:subim1}
\end{subfigure}
\begin{subfigure}{0.325\textwidth}
\includegraphics[width=1\linewidth, height=5.5cm]{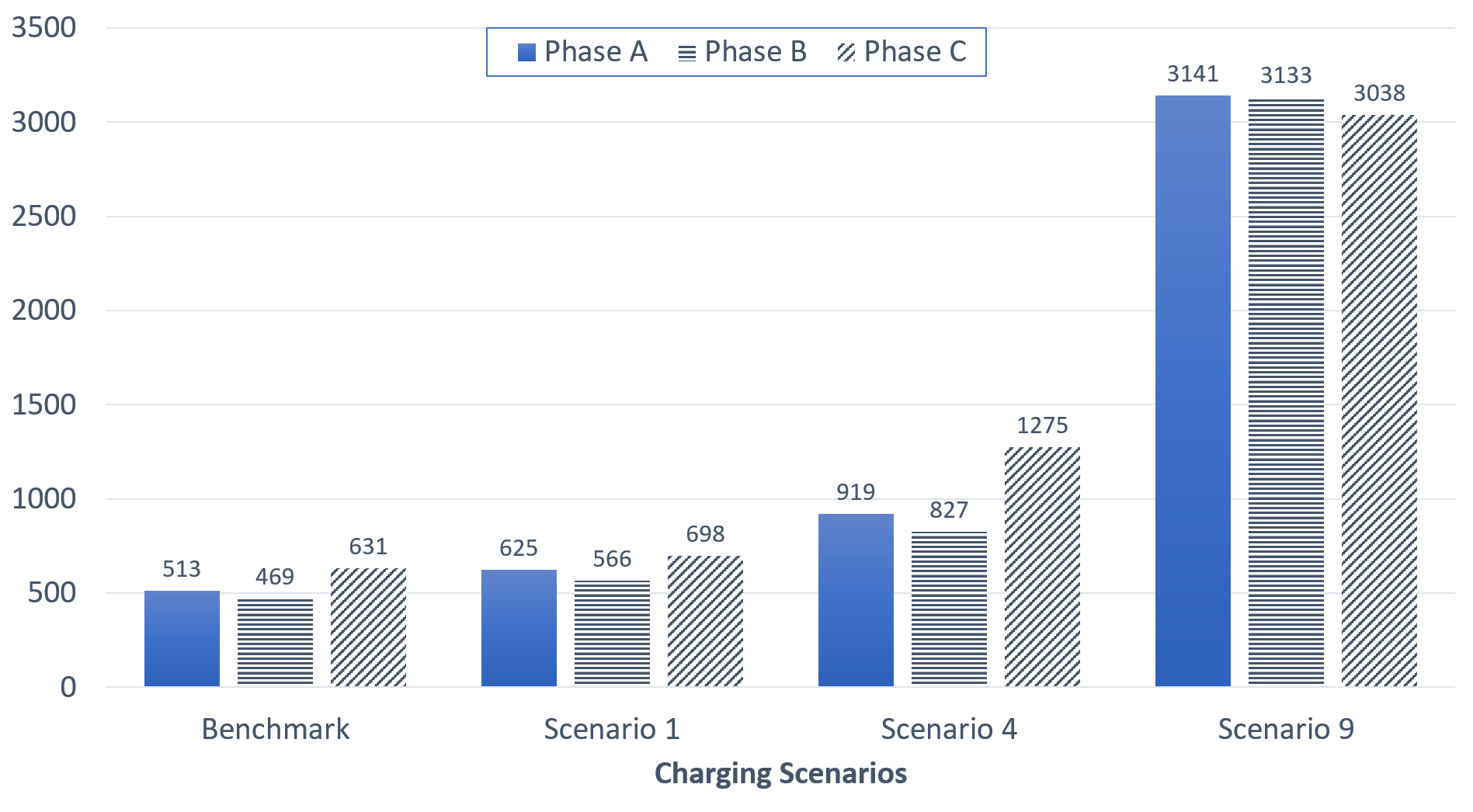}
\caption{Rural Area}
\label{fig:subim3}
\end{subfigure}
\caption{Number of Tap Operations Per Year}
\label{count}
\end{figure}

Fig. \ref{count} shows the total counts of yearly VR tap operations under multiple scenarios for each area with 500 PEV fleets. The inherent unbalanced loading of distribution system caused the different number of tap operations annually for each phase of VR. More importantly, it reveals an insightful observation that the VR operation is highly affected by the impulsiveness of charging activities. Admittedly, Scenario 1 has a higher loading impact due to concentrated home charging actions, however, Scenario 4 has a higher impact on VR operations due to the more frequent charging activities (cf. Fig. \ref{aggregate}). The subsequent LoL of VR can be derived based on \eqref{eq:lvr} and indicates a consistent pattern of lifetime depreciation as in Fig. \ref{count}.

Noted that the two three-phase VRs successively installed at the middle of feeder share the burden of voltage adjusting requirement under highly impulsive PEV load. The more VRs installed in the system, the more controllability of the voltage profile is available in real time, but meanwhile induce more capital investment. The utility is facing more challenges with the current trend of increasing PEV penetration when dealing with such trade-off \cite{yuan2017modified,epri2012understanding}.

\subsection{Effect of Demographic Discrepancy on Asset Depreciation}
All previous case studies exclude the demographic information embedded in different areas, and only consider one charging scenario at each simulation. This section serves as a separate case study to particularly analyze the demographic impact on grid assets, where we have mixed charging scenario at each simulation. The comparing metrics have been modified as follows. For each area, the PEV penetration level (PL) is defined as the ratio of coincidental PEV charging load to the total base load.
\begin{align} 
    PL=\frac{\sum P^{PEV}}{P^{b}}\times 100 \label{PL}
\end{align}

The three areas have different percent composition of slow/fast-charging PEV based on demographic nature, as shown in Table \ref{t6}. Noted that in this study we assume that the percentage of fast charging (120 kW) PEV is in descending order of suburban (highest), urban, and rural area (lowest). This assumption is justified by the fact that the suburban area has the highest possibility for PEV to utilize en route fast charging facility.\footnote{For example, the only Tesla Supercharging Station currently built in Columbus is not located in downtown (urban area), but in Grove City (suburban area).} Hence, it is expected that the power grid in suburban area will be exposed to the most impulsive and drastic PEV charging load, which induces the most depreciation and long-term cost for the grid asset. To observe progressive impact, multiple $PL$ scenarios, i.e., benchmark (no PEV), $50\%$, $100\%$, $200\%$ and $300\%$, are investigated. Same reasoning mentioned in Section \ref{subsec3.2}, the setup and upper bound of $PL$ is set based on the scenario that every household owns a PEV, which is predicted realistic in the future. Noted that the paper does not intend to draw an exhaustive conclusion for various charging and penetration cases, but rather, to propose an integrated algorithm that helps utility interpret PEV's impact. Based on \eqref{PL} and Table \ref{t6}, the detailed number of PEV fleets for slow-charging and fast-charging can be determined respectively, which in total make up the $PL$ of interest.

\begin{table}[!h]
\begin{center}
\caption{Percent Composition of Two Charging Levels} \label{t6}
\begin{tabular}{|c c c|} 
\hline
 & Slow-charging & Fast-charging \\ 
\hline
Suburban Area & 60\% & 40\% \\ 
\hline
Urban Area & 70\% & 30\% \\
\hline
Rural Area & 80\% & 20\% \\
\hline
\end{tabular}
\end{center}
\end{table}

The long-term cost of transformer is estimated with the modified method described in Section \ref{subsubsec_tco} and compared with conventional TCO formulation described in Section \ref{subsec1}. The results are compared over $T_{ins}$. If the transformer is exhausted at $T_x$ before $T_{ins}$ due to the extra stress imposed by PEV loads, then a new transformer is purchased and its induced cost (comprised of capital cost and operating cost) is added to the total cost. The parameters associated with TCO evaluation of the substation transformer are obtained from an anonymous vendor and \cite{rural2016guide}, as summarized in Table \ref{t5}.

\begin{figure}[ht]
	\centering
	\includegraphics[width=1\linewidth, height=7cm]{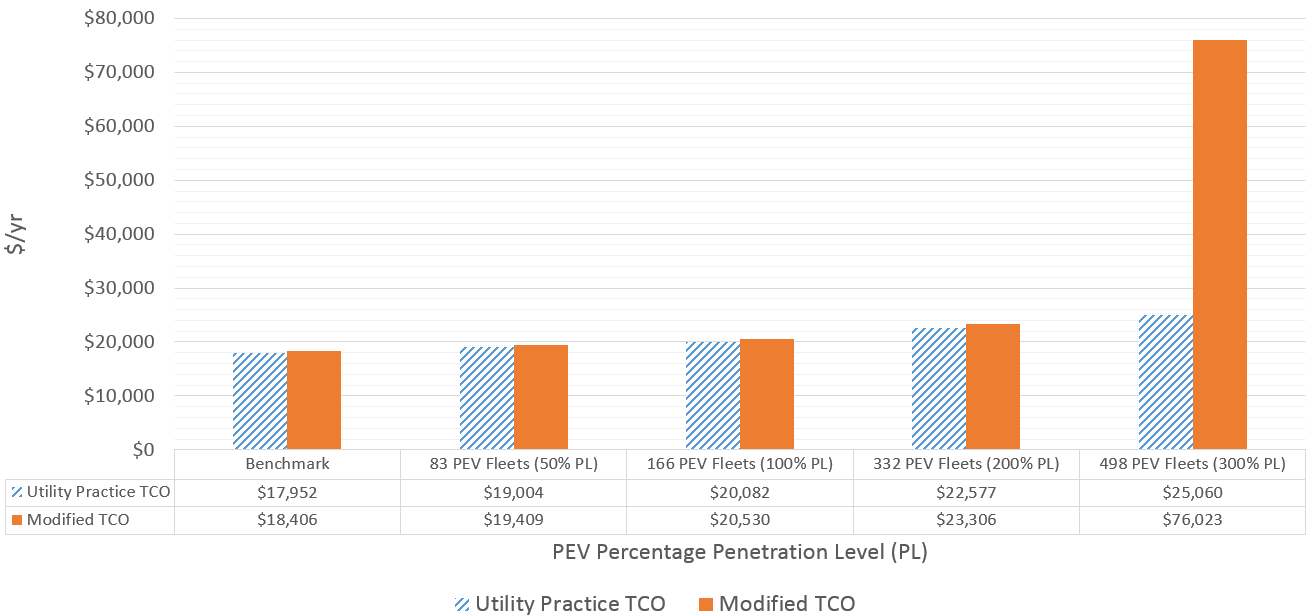}
	\caption{TCO of Transformer in Suburban Area}
	\label{TCO1}
\end{figure}

The TCO of transformer in suburban area estimated with two methods are shown in Fig. \ref{TCO1}. It can be seen that the results of both methods indicate that the long-term cost of the transformer is greatly increased with increasing $PL$. Moreover, results are very close at low $PL$, when $PL$ is greater than $200\%$, the proposed method assesses much higher long-term cost than conventional TCO. This difference in trend is attributed to the fact that substation transformers are usually over-sized for reliability concerns. Therefore, a relatively low $PL$ is not likely to cause a noticeable adverse impact on transformer operation. However, when the grid hosts more PEV, the impact can only be captured accurately with the proposed TCO method. 
\begin{table}[h!]
\centering
\caption{TCO Parameters and Specifications} \label{t5}
 \begin{tabular}{|c|c|} 
 \hline
 Parameters & Value \\ 
 \hline
 $s_R$ [MVA] & 10  \\
 
 $CL$ [kW] & 13.2  \\
 
 $LL$ [kW] & 53  \\
 
 $DC$ [\$/kW-yr] & 120 \\
 
 $RF$ & 0.81 \\
 
 $EC$ [\$/kWh] & 0.05  \\
 \hline
  $\gamma$  & 0.2  \\
 
  $i$ [\%] & 5  \\
 
  $C_o$ [\$] & 70,000  \\ 
 
  Evaluation Period [yr] & 15.41  \\ 
 \hline
\end{tabular}
\end{table}

In this case, the PEV $PL$ is based on a proportion of electric load in each area, which gives us a relatively conservative estimation of PEV load impact. For example, under 300\% $PL$, the suburban area will have 498 fleets in total, which could very well be an understatement considering the case study results in Section \ref{subsec_grid}. But still, we can observe a $\sim 5\%$ accuracy improvement of cost estimation as compared to current utility practice for the lower $PL$ case and $30\% \sim 40\%$ accuracy improvement for the more drastic $PL$ case.

A further question to ask is whether it is more reasonable to use a larger size transformer under high $PL$, which will essentially bring the results of the two methods to the same values. The answer could be case dependent. For example, sometimes a larger transformer could cost more than replacing a small transformer after its end of life, while other times the reverse is true. Nevertheless, even if the planning strategy might conceal the inaccuracy of the conventional TCO method, the fidelity of the proposed method is demonstrated at every $PL$. Moreover, the proposed method enables evaluation of equipment long-term cost over any time span of interest, which provides great flexibility to utilities planning work.
\section{Conclusion}
\label{sec4}
With constantly increasing PEV penetration and improving fast charging technologies, it is critical for utilities to quantify the impact of PEV loads on grid assets and plan for equipment replacement and infrastructure expansion accordingly to ensure service reliability. The unique impulsive characteristics of PEV loads make conventional assessment methods of load impact unsuitable. To address this challenge, this paper proposes an algorithm for evaluating grid assets depreciation under high penetration of PEVs. Compared to the existing evaluation methods, which are case-specific or static, the proposed algorithm provides convenient assessment through an integrated interface and is capable of capturing the inter-temporal response of grid assets. The advantageous features of the proposed algorithm are realized under a mathematical framework, where grid assets' generic models are established and their cost functions are reformulated. In addition, TS analysis and MCS are deployed to ensure the algorithm's accurate and robust performance by accounting for the random charging patterns over time and space. The fidelity of the proposed method is demonstrated on a set of power distribution networks in Columbus metropolitan area, Ohio. The results of this paper can be developed software planning tools for utilities.
\section*{Acknowledgment}
This work was sponsored by the Ford Motor Company. Meanwhile, the authors would like to thank American Electric Power (AEP) for the provision of large-scale electric network data in Columbus metropolitan area.





\bibliographystyle{IEEEtran}

\bibliography{reference}

\end{document}